\documentstyle[twoside,fleqn,espcrc2]{article}

\input{prepictex.tex}
\input{pictex.tex}
\input{postpictex.tex}

\newcommand{\be}{\begin{equation}}
\newcommand{\ee}{\end{equation}}
\newcommand{\bea}{\begin{eqnarray}}
\newcommand{\eea}{\end{eqnarray}}
\newcommand{\CA}{{\cal A}}
\newcommand{\CG}{{\cal G}}
\newcommand{\FP}{\mbox{FP}}
\newcommand{\intl}{\int\limits_{-L}^{L}}

\newcommand{\AmS}{{\protect\the\textfont2
  A\kern-.1667em\lower.5ex\hbox{M}\kern-.125emS}}

\title{Hamiltonian Approach to the Gribov Problem}

\author{T.~Heinzl\thanks{e-mail:
thomas.heinzl@physik.uni-regensburg.de}  \address{Institut  f\"ur
Theoretische    Physik,   Universit\"at    Regensburg,    D-93040
Regensburg, Germany} }

\begin{document}

\begin{abstract}
We study the Gribov problem within a Hamiltonian  formulation  of
pure Yang-Mills theory.  For a particular  gauge fixing, a finite
volume  modification  of  the  axial  gauge,  we  find  an  exact
characterization   of  the  space  of  gauge-inequivalent   field
configurations. 
\end{abstract}

\maketitle

\section{Introduction}

It is well known that the main problem for a quantum  formulation
of gauge  theories  is the fact  that  the space  $\CA$  of gauge
potentials  is `too  large';  there  are  infinitely  many  gauge
equivalent  configurations  corresponding  to one  and  the  same
physical situation.  This huge redundancy should be eliminated by
constructing  the  physical  configuration  space,  the space  of
orbits,  $\CA_{phys}  = \CA / \CG$,  consisting  of gauge  fields
modulo  gauge  transformations,  the elements  of the gauge group
$\CG$.

The   aim  of  this   contribution   is  to  find   the  physical
configuration space $\CA_{phys}$ by identifying a subset of $\CA$
with  $\CA_{phys}$.   The method  for doing  that is the familiar
procedure  of {\em gauge  fixing},  which amounts  to choosing  a
representative  gauge  potential  $A$ on each of the orbits.   In
doing that, two criteria have to be met.

(i) {\em existence:} A representative  should be selected on {\em
any} orbit; no orbit should be omitted.

(ii)   {\em  uniqueness:}   There   should   be  only  {\em  one}
representative obeying the gauge condition on each orbit.  If, on
the other hand, there are (at least) two gauge equivalent fields,
$A_1$,  $A_2$, satisfying  the gauge condition,  the gauge is not
completely  fixed.  There is a residual  gauge freedom given by a
gauge  transformation,  say,  $U$,  between  the `Gribov  copies'
\cite{Gri78}, $U$: $A_1 \to A_2$.

In view  of that,  it is clear  that  the physical  configuration
space $\CA_{phys}$  is the maximal subset of $\CA$ containing  no
Gribov  copies.   It  is called  a `fundamental  modular  domain'
\cite{vBa92}.

In order to find the physical  configuration  space we follow the
pioneering work of Feynman on (2+1)-dimensional Yang-Mills theory
\cite{Fey81}  and make use of the intuitive Hamiltonian formalism
within  a  functional  Schr\"odinger  picture.   As  reemphasized
recently  in \cite{KN96},  there are reasons to believe  that the
fundamental  domain  (at  least  in 2+1 dimensions)  has a finite
volume leading to a purely discrete spectrum.  This would provide
a natural explanation  of a mass gap in the theory.   The quantum
mechanical  analogue  for this  is the infinite  square  well  of
extension  $d$.  The gap between  the ground  state and the first
excited  state  is of the order  $1/d^2$.   The non-abelian  case
corresponds  to finite $d$ and thus to a finite  gap; the abelian
case corresponds  to infinite  size, $d \to \infty$,  and thus to
vanishing gap (the analogue of the massless photon).

\section{An Example from Quantum Mechanics} 

To gain  some  more intuition  let us stay  a little  bit further
within  the  context  of  quantum  mechanics.  Consider  a  point
particle moving in a plane described  by (cartesian)  coordinates
$q_1$,  $q_2$,  and  let  the  particle  have  vanishing  angular
momentum, $G \equiv q_1 p_2 - q_2 p_1 = 0$.  This latter identity
we interpret  as Gauss's  law,  so that  $G$ is the generator  of
gauge transformations which are just rotations around the origin.
If we introduce polar coordinates,  the radius $r$, and the angle
$\phi$, it is obvious that the angle changes under rotations  and
is thus  gauge  variant,  whereas  the  radius  $r$ is the  gauge
invariant variable. Accordingly, the physical configuration space
is the positive real line. The associated physical Hamiltonian is

\be
H_{phys}  = - \frac{1}{2}  r^{-1}  \frac{\partial}{\partial  r} r
\frac{\partial}{\partial r} \; ,
\label{HPHYS}
\ee

\noindent
and depends  only  on the gauge  invariant  variable  $r$,  as it
should.

Let us assume  now that we are not as smart as to guess the gauge
invariant  variables  and proceed  in a pedestrian's  manner  via
gauge fixing.  We gauge away $q_2$, $\chi(q) \equiv q_2 = 0$, and
immediately  realize  that this gauge  choice  selects  {\em two}
representatives on each orbit at $\pm q_1$ (see Fig.~1). There is
a (discrete)  residual gauge freedom between the copies, $q_1 \to
- q_1$.  So we have a Gribov problem,  which is best analysed  in
terms of the Faddeev-Popov (FP) operator \cite{FP67} given by the
commutator of the gauge fixing with Gauss's law,

\be
\FP \equiv -i \, [\chi, G] = q_1 \; .
\ee

\noindent
The latter turns out to be coordinate  dependent  and vanishes at
the  `Gribov  horizon',  $q_1  = 0$,  which  is  just  the  point
separating  the two gauge equivalent regions $q_1 > 0$ and $q_1 <
0$. So we can fix the gauge completely by demanding that $q_1$ be
positive (an inequality,  thus a non-holonomic  constraint),  and
thus again we find that the physical  configuration  space is the
positive  real line where  we can identify  $q_1$ with the radius
$r$.  Due to the simplicity  of the example the decomposition  of
the configuration space into its redundant and physical parts via
gauge fixing can easily be visualized (Fig.~1).

\vspace{.5cm}

\beginpicture
\setcoordinatesystem units <1cm,1cm>
\setplotarea x from -3.5 to 3.5, y from -3.5 to 3.5
\arrow <5pt> [.15,.3] from -3.5 0 to 3.5 0
\arrow <5pt> [.15,.3]  from 0 -3.5 to 0 3.5
\put {$q_1$} at 3.4 -0.3
\put {$q_2$} at 0.3 3.4
\circulararc 360 degrees from 1 0 center at 0 0
\circulararc 360 degrees from 2 0 center at 0 0
\circulararc 360 degrees from 3 0 center at 0 0
\put {$\bullet$} at -3 0
\put {$\bullet$} at -2 0
\put {$\bullet$} at -1 0
\put {$\bullet$} at  1 0
\put {$\bullet$} at  2 0
\put {$\bullet$} at  3 0
\arrow <5pt> [.15,.3] from 1.5 0.35 to 1.5 0
\put {\footnotesize FMD} at 1.4 0.55
\arrow <5pt> [.15,.3] from -2.5 0.35 to -2.5 0
\put{\footnotesize GF} at -2.5 .55
\arrow <5pt> [.15,.3] from 2.83 -2.83 to 2.12 -2.12
\put {orbits} at 2.8 -3.2
\put{
\begin{picture}(2,2)
\setlength{\unitlength}{1cm}
\put(0,0){\circle*{0.2}}
\end{picture}} [Bl] at -.11 0
\linethickness=1pt
\putrule from 0 0 to 3.2 0
\endpicture

\vspace{.5cm}

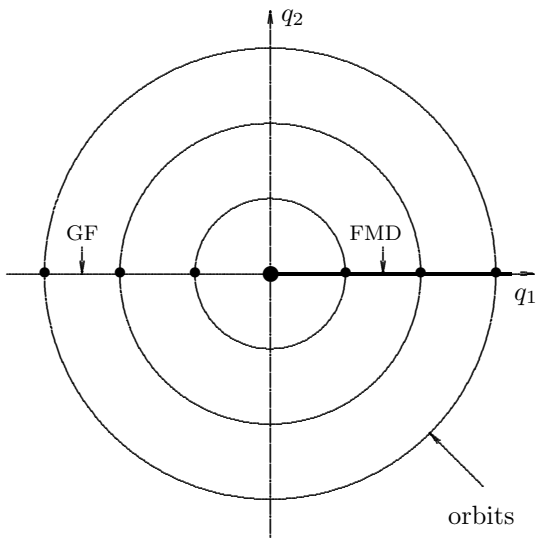
{\sl Figure 1:  The configuration space of the quantum mechanical
example.   The orbits are circles  around the origin.   The gauge
fixing  (GF),  $q_2 = 0$, cuts every  orbit  twice  at the Gribov
copies $\pm q_1$ ($\bullet$).   The Gribov horizon is the origin,
the fundamental modular domain (FMD) the positive real line, $q_1
> 0$.} 

\vspace{.5cm}

Using Gauss's law to eliminate  the unphysical  momentum,  $p_2$,
from the original Hamiltonian,  $H = (p_1^2 + p_2^2)/2$, one ends
up  with  the  physical  Hamiltonian  (\ref{HPHYS}).    Note  the
additional factors of $r = |\det \FP|$ in the kinetic term.  They
guarantee that the Hamiltonian is hermitean on ${\cal L}^2 (r dr,
{\cal R}^+)$.

As a result we can state that gauge fixing amounts to mapping the
original cartesian coordinates  onto curvilinear ones and setting
the gauge  variant  part of these  (the `angles')  to zero.   The
Jacobian of the transformation  is the FP determinant  and enters
the  physical   Hamiltonian   as  well  as  the  scalar   product
\cite{CL80}.

\section{Yang-Mills Theory in Palumbo Gauge}

Let us now consider the case of field theory. According to Dirac,
there is still another,  more physical,  interpretation  of gauge
fixing  as a `dressing'  of the  fermions  \cite{Dir55}.   In the
abelian case, one has the dressed electron

\be
\psi_\chi ({\bf x}) = \psi ({\bf x}) \exp ie \!  \int \! d^3 y \,
{\bf E}_\chi ({\bf x} - {\bf y}) \cdot {\bf A} ({\bf y}) \; , 
\ee

\noindent
where ${\bf E}_\chi$ is the dressing electric field corresponding
to the gauge fixing  $\chi = 0$.  For the Coulomb  gauge it is of
course  the familiar  Coulomb  field \cite{Dir55},  for the axial
gauge, $\chi \equiv A_3 = 0$, it is the singular configuration

\be
E_1 = E_2 = 0\; , \quad E_3 = e \theta  (x_3) \delta (x_1) \delta
(x_2) \, , 
\label{ESTRING}
\ee

\noindent
corresponding  to a string  of electric  flux  in the 3-direction
\cite{Dir55}.   It is easy  to check  that  (\ref{ESTRING})  is a
singular  solution  of Gauss's  law  for  a point  charge  at the
origin,

\be
G \equiv  \partial_i  E_i = e \delta^3  ({\bf  x}) = e \rho_\perp
(x_1 , x_2 ) \rho_L (x_3) \; ,
\ee

\noindent
where the last step stands for a smearing  of the delta functions
in transverse  and longitudinal  direction  with  respect  to the
string.   It has been  known  for a long  time that  the singular
configuration (\ref{ESTRING})  leads to an infrared divergence in
the energy \cite{Dir55} \cite{Sch63} of the form

\be
H \sim e L \left[  \int  dx_3 \, \rho_L  (x_3) \right]^2  + {\rm
finite} \; , 
\ee

\noindent
where we have introduced the length $L$ of the string. The remedy
of this infrared problem is to introduce a homogeneous background
charge  density  in 3-direction,  $\rho_L  \equiv  \delta(x_3)  -
1/2L$. Calculating the integral of the Gauss operator, $G$,

\be
\intl  dx_3  \, G   \sim \intl dx_3 \, \rho_L = 0 \; ,
\ee

\noindent
one finds  that the zero mode of $G$ is the charge  of the string
which  vanishes,  and that  the energy  of the string  is finite.
Thus,  only  neutral  strings  have  finite  energy,  a  property
somewhat  reminiscent  of  confinement.    Due  to  the  manifest
appearance of (chromo-)electric strings, the axial gauge has been
suggested  as  an  appropriate  gauge  choice  for  studying  the
confinement problem \cite{Man79}.

There  are  several   gauges   which  correspond   to  the  above
modification  of  the  axial  gauge  (differing  in the  way  the
residual  gauge freedom is fixed \cite{Hei96}).   However, all of
them have in common that $A_3$ is not completely set to zero, but
only those modes with momentum  $k_3 \ne 0$.  A zero mode, having
$k_3 = 0$, is retained,

\be
A_3 (x_1 , x_2 , x_3 ) \to a_3 (x_1 , x_2 ) \; .
\label{GAUGEFIX}
\ee

\noindent
To fix the residual gauge freedom still left one can do a (self-)
similar construction in 1- and 2-direction.   This is the Palumbo
gauge \cite{Pal86}. Details are unnecessary for what follows.

The  gauge  fixing  (\ref{GAUGEFIX})  can  be shown  to exist  by
explicitly  constructing  the transformation  $U$ that  takes  an
arbitrary configuration  to this gauge.  One finds that $U$ is of
the form $U = W \times V$, with

\bea
W ({\bf x}) &=& P \exp i \int_{-L}^{x_3} dy_3 \, A_3 (x_1 , x_2 ,
y_3 ) \; , \\
V ({\bf x}) &=& \exp [i (x_3 + L) a_3 (x_1 , x_2 )] \; .
\eea

\noindent
The first  exponential,  $W$, takes  one to the pure axial  gauge
($A_3  \to 0$), the second  exponential,  $V$, restores  the zero
mode ($0 \to a_3$), which is determined via $2L \, a_3 (x_1 , x_2
) = \log w (x_1 , x_2 )$ , where $w (x_1 , x_2 ) \equiv  W (x_1 ,
x_2, L)$.  Note that $a_3$ is in the Lie algebra $su(2)$ and thus
obtained  from  the  group  element  $W \in  SU(2)$  by taking  a
logarithm.  As the latter is a multivalued  function we are right
at the question of uniqueness.

The main virtue of the Palumbo gauge (and related  gauges) is the
fact that the FP operator, its inverse and its determinant can be
calculated exactly \cite{Hei96}.   The result for the determinant
is

\be
\det \FP = \frac{\sin^2 |a_3| L}{(|a_3| L)^2} \; ,
\label{DETFP}
\ee

\noindent
with $|a_3| = (a_3^a  a_3^a)^{1/2}$  being the modulus  of $a_3$.
The FP determinant  (\ref{DETFP})  is just  the Haar  measure  of
$SU(2)$, thus the Jacobian of the exponential  map, $\exp:  su(2)
\to  SU(2)$,  from  the  algebra  to the  group.   At the  Gribov
horizons,  where $\det \FP$ vanishes,  one has $|a_3| = \pi k/L$,
$k$ integer,  (in the algebra)  and $w=\pm 1$ (in the group).  At
these points, the exponential  map becomes singular,  the inverse
map,  the logarithm,  becomes  multi-valued  and thus the `angle'
variable $a_3$ ill-defined.  If one writes the zero mode with the
help of a unit vector  $n^a$  in color  space  as $a_3^a  = |a_3|
n^a$,  the  residual  gauge  freedom  can  be  described  in  the
following  way:  one has transformations  $u_1(k)$  which  do not
change  the direction  of the color vector 

\be
u_1 (k) :  |a_3|  n^a \to |a_3 + 2\pi k / L | n^a \; . 
\ee

The  second  type  of  residual  gauge  transformations,   $u_2$,
transforms  along  the  Gribov  horizons  ($w  = \pm 1$)  without
changing the length of the color vector,

\be 
u_2 : |a_3| \, n^a \to |a_3| \, \tilde n^a \; , \quad |a_3| = \pi
k /L \; .
\ee

The fundamental modular domain is therefore given in terms of the
inequality  $| a_3 | < \pi/L$, which describes  the interior of a
sphere in color space~(see Fig.~2).


\beginpicture
\setcoordinatesystem units <1cm,1cm>
\setplotarea x from -3.5 to 3.5, y from -3.5 to 3.5
\arrow <5pt> [.15,.3] from -3.5 0 to 3.5 0
\arrow <5pt> [.15,.3]  from 0 -3.5 to 0 3.5
\put {$a_3^1$} at 3.4 -0.5
\put {$a_3^2$} at 0.5 3.4
\circulararc 360 degrees from 1 0 center at 0 0
\circulararc 360 degrees from 2 0 center at 0 0
\circulararc 360 degrees from 3 0 center at 0 0
\put {$\bullet$} at 0 0
\put {$\bullet$} at -1.4142 1.4142
\arrow <5pt> [.15,.3] from 0 0 to -1.4 1.4
\put {$u_1$} at -1.2 0.7
\put {$\bullet$} at -2.598 -1.5
\put {$\bullet$} at -1.026 -2.819
\circulararc 40 degrees from -2.8 -1.7 center at 0 0
\arrow <5pt> [.15,.3] from -1.15 -3.065 to -1.05 -3.1
\put {$u_2$} at -2.5 -3
\arrow <5pt> [.15,.3] from 2.83 -2.83 to 2.12 -2.12
\put {det FP = 0} at 2.8 -3.2
%
%
\setshadegrid span <2pt>
\setquadratic
\hshade -.707 -.707 -.707 <,z,,>  0 -1 -.707 .707 -.707 -.707 /
\vshade -.707 -.707  .707 <z,z,,> 0 -1     1 .707 -.707  .707 /
\hshade -.707  .707  .707 <z,,,>  0  .707  1 .707  .707  .707 /
\endpicture

\vspace{.5cm}

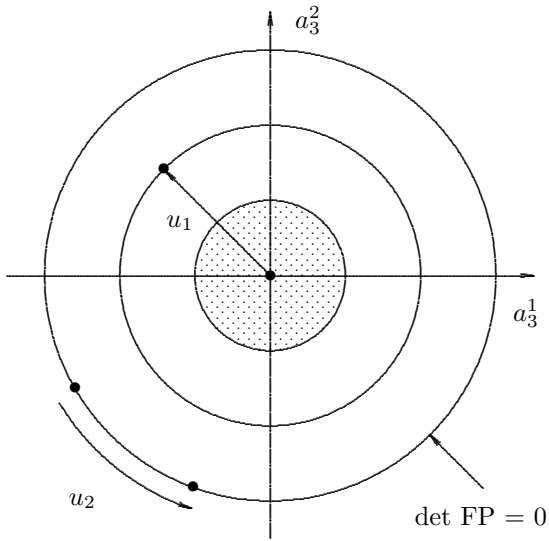
{\sl Figure 2:  A two-dimensional plot of the configuration space
of $a_3^a$,  $a=1,2$.   The Gribov  horizons  where  {\rm det FP}
vanishes  are (apart from the origin)  circles of radius $\pi/L$,
$2\pi/L$ etc. The shaded interior of the inner circle constitutes
the fundamental  modular domain.  The arrows denote the action of
the residual gauge transformations  $u_1 (k=1)$ and $u_2$ between
particular Gribov copies represented by black dots ($\bullet$).}

\vspace{.6cm} 

The physical Hamiltonian containing all the Jacobian factors is a
somewhat lengthy expression and can be found in \cite{Hei96}.

One crucial question remains to be answered:  what is the physics
associated  with the Gribov  horizons  being  discontinuities  in
field space \cite{JMR78}?  To find the answer one would need some
geometrical  space-time  picture  of  the  configurations  at the
horizon.  In related gauges, they seem to correspond to monopoles
\cite{FNP81} or magnetic vortices \cite{Gri95}.   An explicit but
formal construction of a horizon configuration  has been given in
\cite{PS93}.

The center  of the group  seems  to play  a peculiar  role as the
horizon  configurations  correspond  exactly  to  group  elements
valued in the center, $w = \pm1$. Finally, a $\theta$-angle might
emerge  when the wave functional  $\Psi$ `bites  its tail' at the
horizon  with  a  quasi-periodicity   like  $\Psi  (\pi  /  L)  =
e^{i\theta} \Psi (-\pi / L)$.

It might be worthwhile  to study these issues on a lattice  where
one  has  a finite  number  of degrees  of freedom  and  a better
control of infinities.  In the context of maximally abelian gauge
fixing on a lattice,  the Gribov problem and its relation  to the
monopole  condensation  scenario  of  confinement  have  recently
attracted a lot of attention \cite{HT96}.  A lattice gauge fixing
similars   to  ours   has  just   appeared   in  the   literature
\cite{BDH96}.   It looks promising  to investigate  these  issues
further.

\bigskip
\noindent {\bf Acknowledgements}
\smallskip

The author gratefully acknowledges  instructive discussions  with
A.~Jaramillo,  S.~Shabanov  and D.~Zwanziger. It is a pleasure to
thank  the  organizer  S.~Narison  and  his  team  for  all their
efforts.

\end{document}